\documentclass[12pt]{article}
\usepackage{graphicx}
\usepackage{amssymb}
\usepackage{epstopdf}
\newcommand{\ignore}[1]{}
\newcommand{\be}{\begin{equation}} \newcommand{\ee}{\end{equation}}
\newcommand{\ba}{\begin{eqnarray}} \newcommand{\ea}{\end{eqnarray}}
\newcommand{\nn}{\nonumber} \renewcommand{\bf}{\textbf}
\newcommand{\ra}{\rightarrow}

\renewcommand{\a}{\alpha} \renewcommand{\b}{\beta}

\newcommand{\p}{\partial}  
  \newcommand{\NN}{\vec \nabla}

\def\slasha#1{\setbox0=\hbox{$#1$}#1\hskip-\wd0\hbox to\wd0{\hss\sl/\/\hss}}
\def\slashb#1{\setbox0=\hbox{$#1$}#1\hskip-\wd0\dimen0=5pt\advance
       \dimen0 by-\ht0\advance\dimen0 by\dp0\lower0.5\dimen0\hbox
         to\wd0{\hss\sl/\/\hss}}
\def\bra#1{\left< #1\right|}
\def\ket#1{\left| #1\right>}
\def\bracket#1#2{\left<#1\mid #2\right>}
\def\EV#1#2#3{\bra{#1}#2\ket{#3}}
\input{epsf}

\begin{document}

\title{Classifying Polarization Observables \\ of the Cosmic Microwave Background }

\author{John P. Ralston and Pankaj Jain} 

\maketitle

\centering{
 {\it Department of Physics and Astronomy, University of Kansas,\\  Lawrence KS 66045 \\   Physics Department, IIT, Kanpur 208016} }

 \bigskip

\abstract{\small We revisit the classification of polarization observables of the cosmic microwave background. There exists a unified approach to the $3 \times 3$ density matrix by which intensity, linear and circular polarization are treated on an equal footing. The representations of the rotation group on the sphere contain certain right-left setup symmetries which have not been taken into account before.  Left-right symmetries revise the construction of invariants and certain predictions based on symmetries.  There are two true rotational invariants mode by mode in polarization data. Independent of models of cosmological perturbations, we emphasize methods to construct invariant distributions in order to test predictions of isotropy, axial or parity symmetries, and generate covariant and invariant statistics order by order. }

\date{}

\section{Introduction}

Current interest in the cosmic microwave background focuses on testing models of perturbations about an isotropic thermal default.  Polarization data is highlighted for providing new information about the relative size of scalar versus tensor classes of such perturbations.  The groundwork for this task was set up years ago.  And yet, surprisingly, there are gaps in the current approach.  For example,  the distribution of CMB polarizations in various alternative cosmologies might be put into correspondence with the symmetries of the cosmology.  Or one might try to assess deviations from uncorrelated isotropy using empirical distributions.  Quadratic correlations do not encompass such questions. 

Here we revisit the kinematic task of describing polarization and the distribution of polarization, setting aside models of perturbations.  The statistics of linear polarization are more subtle than commonly believed.  The angular orientation of the polarization $\psi$ is measured at polar coordinates $\theta, \, \phi$. Other variables that can be measured are the intensity (commonly called ``temperature''), circular polarization, and the degree of polarization.  Although the different variables have different transformation properties, we set up a framework in which all can be treated together under one unified description.  

The observables of polarized light are the wavenumber $\vec k =k \hat k$ and polarization density (coherency) matrix $\rho_{ij} = <E_{i} E_{j}* >$ for 3-component electric field $E_{i}$.  Since the field is transverse, it is conventional to reduce it to the transverse components, $\rho_{ij}^{\perp}= <E_{i}^{\perp}E_{j}^{* \perp}>.$  This step complicates the full classification, and we do not use it.  Next the astronomical conventions \footnote{Actually astronomers use $RA$ and $DEC$.} employ unit vectors $\hat \theta, \, \hat \phi$ in the tangent plane to the sky, introducing a natural basis but also a troublesome coordinate-dependence. The $2 \times 2$ Hermitian matrix is expanded in a local complete set of tensors, via Stokes parameters:  
\ba    \rho^{\perp}= {1 \over 2 }I \delta_{ij}^{\perp} + {1 \over 2 }U 
t_{ij}^{B} + {1 \over 2 }Q t_{ij}^{E} + {i \over 2} t_{ij}^{V}; \nn \\     t_{ij}^{E} =  \hat \theta_{i} \hat \theta_{j}- \hat \phi_{i}\hat \phi_{j}; \nn \\    t_{ij}^{B} =  \hat \theta_{i}  \hat \phi_{j} + \hat \theta_{j}\hat \phi_{i}; \nn \\ t_{ij}^{V}=   \hat \theta_{i} \hat \phi_{j}- \hat \phi_{i} \hat \theta _{j}.  \ea Here $\delta_{ij}^{\perp}$ is the 2 $\times $2 unit matrix orthogonal to the direction of observation $\hat r$.  By construction here, \ba  Q = {1 \over 2}{  tr(\rho t^{E}) \over \sqrt{tr(\rho)} } ; \nn \\ U = {1 \over 2} {  tr(\rho t^{B}) \over \sqrt{ tr(\rho)}  }, \label{elecmag}  \ea where $tr$ indicates the trace.   Born and Wolf use $s_{0}=I, \, \vec s= s_{j}= Q, U, V$. A pure polarization state oriented at angle $\psi$ relative to $\hat \phi$ has $Q= cos 2 \psi$, $U= sin 2 \psi$.   Under a rotation by angle $\delta$ about the line of observation $\hat r$, $Q$ and $U$ mix:  \ba   \left(\begin{array}{c} Q \\ U \end{array}\right)     &\ra &  \left(\begin{array}{cc}  cos 2\delta & sin 2\delta \\ -sin 2\delta &  cos 2\delta \end{array}\right)  \left(\begin{array}{c} Q \\ U \end{array}\right) ; \label{EBmix} \\ \psi &\ra &  \ \psi +2 \delta  \ea The transformation property via twice the angle represents the tensor nature of the density matrix. 

Given some convention for interesting modes, one might predict them in a model for perturbations in the $\hat \theta$, $\hat \phi$ frame. The observer can then measure the modes in just the same coordinate frame to confront the model.  Both model and observations then contain coordinate dependence, which stand to complicate or invalidate model-independent tests and tests based on symmetry.  Towards dealing with these problems, Zaldarriaga and Seljak (ZS)\cite{ZS} and Kosowsky, Kamionkowsy and Stebbins (KKS)\cite{KKS} developed elegant rules for projecting $\rho^{E}$ and $\rho^{B}$ into scalar fields.  Recognizing that the linear polarizations in $\rho^{\perp}$ are linear combinations of helicity $\pm 2$, ZS used the technology of Newman and Penrose\cite{NP,Goldberg:1966uu} called ``spin-lowering'' to cast the observables into invariants.  KKS project the 2 $\times $ 2 density matrix into a fiducial basis more directly, and show their method is equivalent.  Our description avoids spin-lowering and the entire 2 $\times $ 2 approach for technical reasons we will review.  

\subsubsection{An Instructive Paradox}

The need for an appropriate framework for statistics of angular variables is illustrated by an instructive paradox from biology\cite{Batschelet}.  An observer studying migrations records angles of bird flights $\theta_{i}$ relative to East.  The distribution is actually isotropic, as (say) 360 birds fly in 360 different directions separated by one degree. The observer makes a distribution $f(\theta)=dN/d\theta$ from the data.  Classical statistics suggests the average angle $<\theta>=\sum_{i}^{N} \theta_{i}/N$ is informative.  For the isotropic distribution spread over $0<\theta < 360^{o}$ he finds $<\theta>= 180^{o}$, from which the observer reports that the average bird flies West. 

An obvious error is that $\theta$ is a convention-dependent variable with the unpleasant transformation property $\theta \ra \theta +\delta$ when the coordinate system is rotated by $\delta$.  This problem is addressed by making explicit the coordinate origin with an explicit angle $\theta_{East}$.  Then $\theta -\theta_{East}$ is invariant under change of origin.  All higher moments of $(\theta -\theta_{East})^{A}$ are invariant.  Yet despite invariance, such moments {\it generally continue to depend upon the convention} for $\theta_{East}$! 
The reader can verify this with a few numerical integrals in the prototype distribution $exp(- k cos(\theta-\theta_{East}))$.  For $k=1$ the values $(\theta -\theta_{East})^{2}$ go from about 5.2 to 11.5, depending on $\theta_{0}$. 
Here $\theta_0$ is an arbitrary parameter defining the zero of the coordinate.
We call moments of this kind ``convention dependent invariants'' ($cdi$): they typify the troubles of circular statistics. 

Towards correcting such diseases one first decides upon a symmetry group of transformations that will be used.  One then maps the data into representations.  Group representations are complete set of elements that transform linearly.  Coordinate changes and the construction of meaningful invariants are greatly expedited.  Yet the bird-flight example reminds us that {\it invariance is necessary but not always sufficient for data analysis to be independent of conventions.}  Non-Abelian groups, in particular, have coordinate conventions that depend on {\it ordering}. 

For polarizations the $3 \times 3$ density matrix at each fixed position $\hat r$ is a reducible tensor of rank 2 under the spin rotation group, which 
decomposes $1\times 1 \ra 2+1+0$, where the integers label the spin. Spin representations are labeled by a second quantum number $s$, thus $$ \rho_{S, \, s} =\Gamma_{ij}^{S, \, s} \rho_{ij}, $$ where $\Gamma_{ij}^{S, \, s} $ is a Clebsch matrix that depends on frame conventions.  Then $\rho_{S,s}(\theta, \, \phi)$ 
over the dome of the sky has an expansion \ba \rho_{S,\, s}(\theta, \, \phi)= \rho_{S, \, s, \,m}^{j} Y_{m}^{j}(\theta, \, \phi), \ea where $Y_{m}^{j}$ are the spherical harmonics. By construction these $\rho_{S, \, s , \, m}^{j}$ elements transform like a tensor under a $SO_{spin}(3) \times SO_{orbital}(3)$ group of 6 parameters acting on the indices $m$ and $s$, respectively.   

The actual continuous transformation group for rotations is $SO_{total}(3)$, whereby the 6 possible parameters of $SO_{spin}(3) \times SO_{orbital}(3)$ are reduced to 3 common parameters used in one joint rotation.  There remains 3 fixed parameters associated with the ordering of operations.  There is hidden freedom in how the groups are joined together.  Extra work, described below, is needed to cast $\rho_{ij}(\theta, \, \phi)$ into representations of $SO_{total}(3)$ taking into account all parameters.  Our task is to explain and explore the consequences of the extra freedoms for observables versus $cdi$. 

There is an interesting parallel with Wigner rotations known in defining spin-states within representations of the Lorentz group.  More than one Wigner rotation can be used to make a basis!  After a good basis is built, there comes a step of {\it classification} of its elements under other transformations such as parity.   Good classification reveals {\it convention dependence among invariants}, which will otherwise hide if standards end at mere invariance. One can then categorize {\it observables} hinging on invariant comparisons of two or more measurements.  Finally, classification of the observables leads to classification of the {\it distributions} of the observables which allows symmetries to be tested.  

Our focus, then, is on making invariant distributions and using them to probe the symmetries of the CMB in a model-independent way.  We pay careful attention to exposing human conventions that cannot be a feature of Nature. Yet we remind the reader in advance of basic validity in setting up coordinate systems full of conventions and comparing models with data in just the same conventions. Many possible frameworks may serve so long as conventions drop out between theory and data analysis.  The reverse is not true: not all ``scalars'' constructed in the existing framework are convention-free observables.  For this reason the distinction between an ``invariant'' and an ``observable'' makes a difference depending on how data is analyzed.   

\section{Transformation Properties} 

We start with the $3 \times 3 $ density matrix $\rho(\vec x)$ because it is what is measured. Let us review the transformation properties of such fields under $SO_{total}(3)$. 

Let $\Phi_{\a}^{S}(x)$ be a set of fields transforming under an irreducible representation of spin $S$ of the rotation group. With $R_{\a\b}^{S}$ a rotation matrix, the transformation rule is \ba \Phi_{\a}^{S}(x) \ra \Phi_{\a}^{S'}(x')=R_{\a\b}^{S}\Phi_{\b}^{S}(x). \ea  
Let \ba R_{\a\b}^{S}= \delta_{\a \b}+ \theta^{k}\sigma_{\a\b}^{k}; \nn \\ 
x_{j}' = \delta_{ij}+ \theta_{k}\epsilon_{ijk}x_{j} , \ea where $\theta_{k}$ are the parameters, $\sigma_{\a\b}^{k}$ is the rotation generator for the {\it spin} of the field $\Phi_{\a}$, and $\epsilon_{ijk}$ is the rotation generator on 3-vectors. The $\sigma$ matrices are matrix elements of rotation operators $\vec S$, $\sigma_{\a\b}^{k} =\EV{\a}{S^{k}}{\b}$.   By definition \ba \ [ \sigma^{i}, \, \sigma^{j} \ ] =i \epsilon_{ijk}\sigma^{k},  \ea for $i, j, k, $ cyclic. 

The infinitesimal change $\delta \Phi_{\a}(x') = \Phi_{\a}^{'}(x')-\Phi_{\a} (x)$ is \ba \delta \Phi_{\a}(x) =    \delta \Phi_{\a}^{spin} + \delta \Phi_{\a}^{orbital}; \nn \\ \delta \Phi_{\a}^{spin}=\theta_{k} \sigma_{\a\b}^{k} \Phi_{\b}(x'); \nn \\ \delta \Phi_{\a}^{orbital}=-\theta_{k}\epsilon_{ijk} x_{i}^{'}  {  \p \over \p x'_{j}} \Phi_{\a}(x') .  \ea This yields the orbital generator $$ L^{k} = -i \epsilon_{ijk}  x_{i}^{'}  {  \p \over \p x'_{j}} ,$$ with the algebra   \ba \ [ L^{i}, \, L^{j} \ ] =i \epsilon_{ijk} L^{k},  \ea for $i, j, k, $ cyclic.    The $SO_{spin}(3) \times SO_{orbital}(3)$ have now been exhibited: the final step is to unite them under $SO_{total}(3) $ defined by the diagonal subgroup transforming with the same parameters on both groups:  \ba \vec J = \vec L + \vec S \ea  The point of this review is that rotations employ {\it two a-priori independent transformations} accounting for the change in the spatial location (orbital) and the re-orientation of the indices (spin) implied by a rotation of a tensor field. 

\subsection{Frame Fixing}
\begin{figure}
\begin{center}
\includegraphics[width=4in]{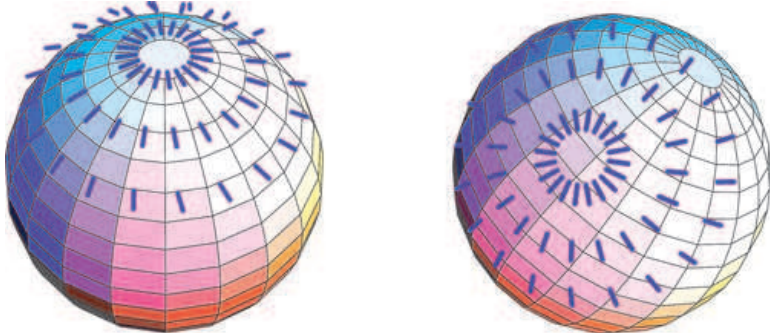}
\caption{\small A simple illustration of frame fixing.  The coordinate lines for the location on the sphere (orbital angular momentum) have no a-priori relation to the abstract frame for spin-coordinates of  ``sticks''.  Although the convention on the right is not standard, it is as valid as any other.  }
\label{fig:CombinedFrames}
\end{center}
\end{figure}

Since the $SO_{spin}(3)$ and $SO_{orbital}(3)$ are independent, there is nothing in mathematics to orient the basis of one relative to another: they as independent as angular momentum and isospin up to this point.  

{\it Phase Conventions and Ordering:}  In beginning quantum mechanics one diagonalizes $S^{z} \ra s_{z}$ and $L^{z} \ra m$.  Choosing the same spin and space frame is a free convention since $ \ [  \vec L, \vec S \ ]=0.$  There remains a need to fix the phase of the eigenstates, which can be interpreted as the orientation of the $x, \, y$ axes on the spin-world relative to the $x', \, y'$ axes on the orbital world.  This small point can be found in the book by Wigner\cite{wigner}: it creates a phase convention on each eigenstate.  Yet since the eigenstates of $J^{z}= L^{z}+ S^{z}$ already have phase conventions, and the rules of quantum mechanics make the conventions unobservable, the relative orientation of space and spin frames is invariably ignored.  But the analogous problem on the sphere will become a central issue. 

For CMB polarizations it is more natural to diagonalize the {\it helicity} of the density matrix, the operator $\hat r \cdot \vec S$. Notice however that $\hat r \cdot \vec S$ does not commute with $L_{z}$: \ba    \ [   L^{i}, \vec S \cdot \hat r \ ]= i \epsilon_{ijk} \vec S^{j}\hat r^{k}. \ea Thus if we operate with $\hat r \cdot \vec S$, followed by a rotation of the space frame, it is not the same as rotating the space frame followed by operating with $\hat r \cdot \vec S$.  Nevertheless  $\vec S \cdot \hat r$ is a global scalar under the {\it total} rotation group,  \ba \ [   J^{i}, \vec S \cdot \hat r \ ]=0. \ea This causes eigenstates of $\hat r \cdot \vec S$ to be convention-dependent.  In particular, their phases depend on the convention and the order by which they are constructed.  

To appreciate this dilemma, consider two natural procedures to set up coordinates on the sphere.  {\it Procedure 1} starts at the top of the sphere, the $z$ axis, and diagonalizes $L_{z}$ and $\hat r \cdot \vec S=S^{z}$ to make states $\ket{m, \, s}_{1}$. Let the usual $\hat x$ and $\hat y$ be orthogonal unit vectors to complete the definition of the spin basis.  Any relative rotation about to the $z$ axis of the space-and spin-frames leads to a phase which is ignorable.  Define other states $\ket{ \psi, \theta, \phi}_{1}$ by a rotation, \ba \ket{\vec \theta }_{1} =exp( -i \vec J \cdot \vec \theta)  \ket{m, \, s}_{1} \ea  Under the same rotation \ba \vec S \ra \vec S' = R \vec S \nn \\ \hat r \ra \hat r'= R \hat r, \nn \\ \vec S\cdot \hat r \ra \vec S\cdot \hat r . \ea  Thus \ba \vec S\cdot \hat r \ket{\vec \theta }_{1} =s \ket{\vec \theta }_{1}  .  \ea With $exp( -i \vec J \cdot \vec \theta) = R(\psi)R(\theta)R(\phi)$ this procedure is equivalent to parallel transporting the $x, y, z$ basis to the conventional $\hat \theta, \, \hat \phi , \, \hat r$ basis with a final rotation about the $\hat r$ axis by $\hat \psi$. 

Compare {\it Procedure 2}, under which we first rotate the spin basis by arbitrary Euler angles $\a, \beta, \gamma$, thus $ S^{i} \ra R(\a, \beta, \gamma)_{ij}S^{j}.$  Since $\hat S \cdot \hat r$ is a scalar, it is equivalent to rotating $\hat r^{i} \ra  \hat r(\a, \beta, \gamma)= R(\a, \beta, \gamma)_{ij}\hat r^{j}.$  The frame defining orbital states is unchanged.  Diagonalize $\hat r' \cdot \vec S$ to make states $\ket{m, \, s}_{2}$.  Define other states $\ket{ \psi, \theta, \phi}_{2}$ by a rotation, \ba \ket{\vec \theta }_{2} =exp( -i \vec J \cdot \vec \theta)  \ket{m, \, s}_{2} \ea  Under the same rotation \ba \vec S \ra \vec S' =R \vec S \nn \\ \hat r' \ra \hat r''=R \hat r' , \nn \\ \vec S\cdot \hat r' \ra \vec S'\cdot \hat r ''. \ea  Thus \ba \vec S'\cdot \hat r'' \ket{\vec \theta }_{2} =s \ket{\vec \theta }_{2}  .  \ea  This shows there are an infinite number of distinct ways to set up ``a basis'' while the helicity quantum number remains invariant under spin-rotations.  One cannot ask Nature to know which procedure will be used. 

A little thought shows that there is no absolute criterion from mathematics alone to fix such conventions on a sphere; Figure \ref{fig:CombinedFrames} shows a cartoon illustrating the problem.  In much the same way, helicity states in relativistic field theory must be defined by some particular sequence of operations known as {\it Wigner rotations}.  Starting from a particle at rest, a convention must be chosen to define its spin frame. This is followed by a rotation to orient the helicity states, which are finally followed by a boost.  Since boosts and rotations do not commute there is an obvious convention dependence in the basis.  Just as with $SO_{total}(3)$ being made from the direct product of two commuting groups, the Lorentz group can be decomposed into commuting $SU(2) \times SU(2)$ to make representations.  The relative orientation of the separate product group frames is an arbitrary convention, and nothing physical can depend on it.  

\subsection{Left $\otimes$ Right Setup Symmetry}

To build up representations it suffices to take outer products $\ket{s, S} \bra{m, L}$ in all possible combinations.  We used freedom to chose one frame in the dual space, because it shows immediately that there exists unitary representations of the form $U_{sm}, \, UU^{\dagger} = U^{\dagger}U=1,$ with $S=L=j$. These are the representations made famous by the $SO(3)$ left-right sigma model.  They transform under {\it left}-and {\it right}-groups as $$ U \ra V_{L}UV_{R}^{\dagger},  $$ where $V_{L}V_{L}^{\dagger}=V_{R}V_{R}^{\dagger}=1$.  Note that $V_{L}\neq V_{R}$ in general.  The freedom of spin-versus-orbital frame fixing amount to a Left-$\otimes$ Right setup symmetry in defining a basis. 

It is well known that unitary $(j, \, j)$ representations of $SO(3)$ make a complete set for function of three Euler angles. Defining as usual \ba D_{sm}^{j}(\a, \, \b\, \gamma) = \bracket{smj}{U(\a, \, \b\, \gamma)}, \ea the completeness and orthogonality relations are: \ba  \sum_{smj } \,  D_{sm}^{j}(\a, \, \b\, \gamma)D_{sm}^{j*}(\a', \, \b' \, \gamma') &=&    {   8 \pi^{2}    \over 2j+1}     \delta(\a-\a')\delta(cos\b -cos \b') \delta(\gamma-\gamma'), \nn \\    \int d\Omega \,  D_{sm}^{j}(\a, \, \b\, \gamma)D_{s'm'}^{j'*}(\a, \, \b \, \gamma) &=& {   8 \pi^{2}    \over 2j+1}    \delta_{mm'}\delta_{ss'}\delta_{jj'}.  \label{complete}\ea Here $d\Omega = d\psi d \cos \theta d\phi$. Due to Left-Right setup symmetry it is an equally good expansion if we re-orient the spin-frames relative to the space-frames by an arbitrary rotation $R_{ss'}^{j}(\a_{0} \, \b_{0}, \, \gamma_{0})$:  \ba  \tilde D_{sm}^{j}(\psi, \, \theta, \, \phi) = R_{ss'}^{j}(\a_{0}^{j} \, \b_{0}^{j} , \, \gamma_{0}^{j} )D_{sm}^{j}(\psi, \, \theta, \, \phi); \nn \\     \int d\Omega \,  \tilde D_{sm}^{j}(\a, \, \b\, \gamma) \tilde D_{s'm'}^{j'*}(\a, \, \b \, \gamma)= \delta_{mm'}\delta_{ss'}\delta_{jj'}; \nn \\  . \label{complete2}  \ea  We could have also rotated the orbital $m$ indices, of course.   {\it There is one free-rotation for each $j$}, so the freedom is considerable.  

\subsubsection{Criteria for Observables}

\begin{figure}
\begin{center}
\includegraphics[width=3in]{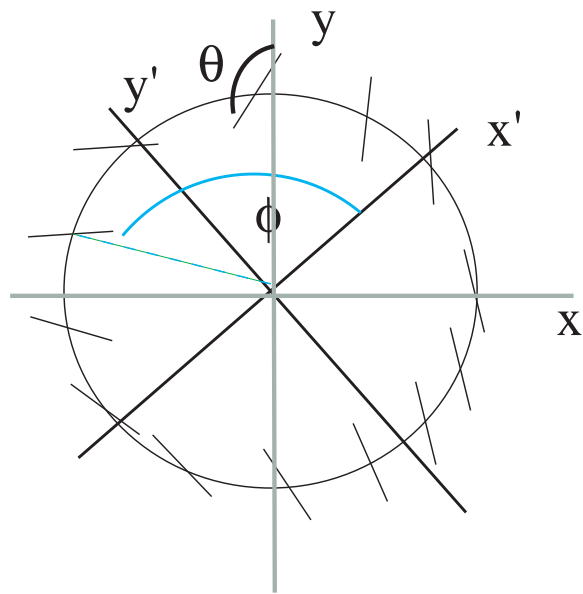}

\caption{Frame-fixing for a direct product of two Abelian groups.  The coordinate frames for position angles $\phi$ versus ``stick angle'' $\theta$ are independent. The facts of Abelian groups (see text) allows the issue to be ignored and the naive power to be a true scalar. 
 }
\label{fig:SO2Frames.eps}
\end{center}
\end{figure}

For reference we define a {\it true scalar} $\phi_{0}$ as a quantity invariant under a one-time Left-Right setup $SO(3) \, R(\a_{0} \, \b_{0}, \, \gamma_{0})$, followed by continuous joint $SO(3)$ $R(\a  \, \b , \, \gamma )$:  \ba \phi_{0} \ra R(\a  \, \b , \, \gamma )R(\a_{0} \, \b_{0}, \, \gamma_{0})\phi_{0}R^{\dagger}(\a  \, \b , \, \gamma ) =\phi_{0}. \ea This remnant of the original $SO(3) \times SO(3)$ transformations reduces 6 continuous parameters to three continuous plus three fixed parameters. 

We define a {\it dependent scalar} as a quantity unchanged when $\a_{0}=\b_{0}= \gamma_{0}=0$ are fixed. Observables that do not depend on the setup of spin and space frames must be true scalars. Dependent scalars can serve for convention-dependent invariants. 

The distinction of true and dependent scalars is a general feature of making product group representations. 
We can illustrate it intuitively with the group $SO(2) \times SO(2)$ 
(Fig. \ref{fig:SO2Frames.eps}).  
Let $\phi$ be the azimuthal angle of a point on a circle in a 2-dimensional plane.   Let $\chi$ be the angle orienting a unit-length ``stick'' glued to each point on the circle - equivalent to a linear polarization.   A continuous function $f(\phi, \, \chi)$ has the conventional expansion \ba f(\phi, \, \chi) =\sum_{mn} \, f_{mn}e^{i m \phi}e^{2 i n \chi}. \ea Each exponential factor is an eigenstate of rotations about the origin in each product space. Since there are two independent origins of the angles, the more explicit expansion including information on angular origins is \ba f(\phi, \, \phi_{0} ;\, \chi,\, \chi_{0}) =\sum_{mn} \, f_{mn}e^{i m (\phi-\phi_{0})}e^{2 i n (\chi-\chi_{0})}. \ea A change in the offset of either origin is the transformation \ba  f_{mn} \ra  f_{mn}e^{-i m(\phi_{0}-\phi_{0}') -2i n( \chi_{0}-\chi_{0}')}. 
\label{off} \ea 
Happily the usual power $|f_{mn}|^{2}$ is a true scalar. In this case it does not matter whether one attends to the difference of partial or true scalars.  Unfortunately this convenience does not extend to non-Abelian groups, where the order of operations determines the properties of the basis elements.  

Yet there is no need to overcomplicate matters.  If one sought general covariance under arbitrary transformations, a coordinate grid where polarizations are measured might be an arbitrary curvilinear system independent of a curvilinear system locating the object positions.  But we are not discussing any transformations more complicated than rotations. It is enough (but necessary!) to deal with the complication of frame fixing.  In the next Section we discuss how the convention dependence originated. 

\subsubsection{Parity Versus CMB $s$-Flip} 

Classification is completed by {\it parity} $P$.  The conventional parity transformation for a tensor operator $t(\vec x)$ of spin $S$ is \ba t(x) \ra Pt(x)P^{\dagger}= (-1)^{S}t(- \vec x).\nn  \ea  Thus $P=P_{S}P(L)$ where $P=P_{S}$ and $P=P_{L}$ are the spin- and space- parity, respectively.  Angular momentum $\vec J$ commutes with parity: \ba  P \vec J P^{\dagger} =\vec J; \label{jtrans} \\  \ [ P ,  \, \vec J \ ]=0. \nn \ea  Hence parity is a scalar operator. Position anti-commutes: \ba P\vec x P^{\dagger} =-\vec x; \label{xtrans} \\ \{ P ,  \, \vec x \}=0. \nn \ea  It is elementary but not uncommon to see the signs of Eq.\ref{jtrans}, \ref{xtrans} reversed, which makes a difference in the following. 

True parity transformations of CMB density matrices relate the data on opposite sides of the sphere.  In the CMB literature the operation of reversing $\vec S \cdot \hat r$ with fixed direction $\hat r$ is often called ``parity,'' but since $\hat r$ reverses under parity (Eq. \ref{xtrans}) we prefer ``CMB $s$-flip'' $\pi_{s}$.  Then \ba \pi_{s} \vec S \cdot \hat r\pi_{s} =-\vec S \cdot \hat r; \label{hel1} \\ \pi_{s} \vec L \cdot \hat r\pi_{s} =+\vec L \cdot \hat r, \label{hel2} \ea The algebra implies that quantum numbers $s \ra -s, \, m \ra m$: \ba  \vec S \cdot \hat r \pi_{s} \ket{s, \, m} = - \pi_{s}\vec S \cdot \hat r \ket{s, \, m}=-s \pi_{s}\ket{s, \, m}.\label{algy} \ea  One defines $E$ and $B$ modes as even or odd under $s$-flip: \ba  D_{m}^{Ej} =  -(D_{2m}^{j} + D_{-2m}^{j})/2 ; \nn \\  D_{m}^{Bj} =i( D_{2m}^{j} -D_{-2m}^{j})/2; \nn \\ \pi_{s}D^{E/B} = \pm D^{E/B}, \ea where the slash indicates the $E$ or $B$ case. 

Now as a consequence of Eqs. \ref{hel1}, \ref{hel2}, the CMB $s$-flip is {\it not} a scalar under rotations.  Since the $E$ and $B$ modes are made from spin-frames of a particular convention, they retain the convention.  This is intuitively apparenty if one temporarily forgets about the global pattern of $E$ and $B$ mode ``cartoons,'' and simply looks along a line of sight in some fixed $\hat r$ direction. Locally one sees a single stick whose orientation is pure convention relative to an arbitrary local coordinate system.  The data is local, but the classification is not!  The transformation of $E$ and $B$ modes into one another under spin-rotations is simply Eq. \ref{EBmix} once again.  

There are several reasons for missing this simple transformation.  First, {\it true parity} is a true scalar operator that is very familiar, and classification by parity cannot change under full rotations.  By replacing parity with $\pi_{s}$ one tends to take true invariance for granted. Second, conversion of $D_{sm} \ra D_{m}^{E}, \, D_{m}^{B}$ eliminates the $s$ index which transforms, creating a notation with appearance of true invariance.  It is illusory because the phase of the $\vec S \cdot \hat r$ states is free, which is just the same phase that varies under under spin-rotations, and just the same phase fixed by hidden conventions.  Earlier we mentioned that $-\vec S \cdot \hat r $ is {\it not} a true scalar under either the spin or orbital groups, but only jointly.  Classifying $E$ and $B$ using spin-parity relative to the spin-group causes $\pi_{s}$ not to be invariant under Left-Right setup symmetry.

Finally it is easy to confuse CMB $s$-flip with spin-parity $P_{s}$ and orbital parity $P_{L}$.  There exists a notion of ``CMB $m$-flip'' $\pi_{m}$ completing the algebra: \ba \pi_{m} \vec S \cdot \hat r \pi_{m} = +\vec S \cdot \hat r; \label{hel3} \\ \pi_{s} \vec L \cdot \hat r\pi_{s} =-\vec L \cdot \hat r, \label{hel4} \ea Note the signs.  As with eq. \ref{algy} this algebra implies that quantum numbers $s \ra  s, \, m \ra -m$. Overall parity $P = P_{S}P_{L}= \pi_{m} \pi_{s}$, by which the product makes the usual scalar operator.  

Why are $E$ modes and $B$ modes nonetheless thought to be invariant classes?  It is because they are {\it dependent invariant classes}, invariant only under fixed spin-frame and orbital-frame transformations, not fully invariant classes. 

Why don't these intricacies plague conventional classifications?  It is because spin-parity and orbital parity are true scalars, {\it not} referring to a spin-frame convention.

\section{Background Review} 
Here we review background related to the frames and their history.

\subsection{Euler Angles and $D$ Matrices}

The simple and natural coordinate system used in astronomy\footnote{Sign conventions of course must be taken into account} happens to correspond beautifully with the formal procedures used in defining our basis. 

Take the $XYZ$ coordinate frame with origin $O$ and translate it without rotation to the top of a unit sphere.  Rotate by angle $\phi$ about $OZ$, so that the new unit vector $\hat x$ is locally parallel to $\hat \theta$. Rotate by angle $\theta$ around $Y'$ in this system, sending the axis $Z \ra Z'$ located along $\theta, \, \phi$.  Rotate around the new $Z'$ axis by $\psi$ until it is parallel to the linear polarization.  The sequence of rotations is denoted by the operator $R(\phi, \, \theta,\, \psi)$.  It is easily shown\cite{sakurai} to be equivalent to first rotating by $\psi$ about $OZ$, then $\theta$ about $OY$, then $\phi$ about $OZ$.   This allows identification of the matrix elements, and in particular diagonalization about the fixed axis $OZ$, with the result \ba D_{sm}^{j}(\phi, \, \theta, \, \psi) = e^{is \psi}d_{sm}^{j}(\theta)e^{im \phi}. \nn \ea The $d_{sm}^{j}$ matrix elements are attributed to Wigner, who developed an explicit formula: \ba  D_{sm}^{j}(\phi, \, \theta, \, \psi) &=&    \ [ {(j+m)!(j-m)! \over (j+s)! (j-s)! } \ ]^{1/2}\nn \\ & \times &  \sum_{r}    (-1)^{j+s-r}    \left(\begin{array}{c}j+s \\r\end{array}\right)\left(\begin{array}{c}j-s \\r-s-m\end{array}\right) \nn \\ & \times & e^{i s \psi}sin^{2j}(\theta/2) cot^{2 r-m-s} (\theta/2   )   e^{im \phi} .\nn \ea 

For practical data analysis it is straightforward to enter the $D_{sm}^{j}$ formula into a computer once, transfer the angles $\phi, \, \theta, \, \psi$ for each data point, and generate a corresponding array of representations for each $j$. 
  
\subsubsection{Spin-Lowering} 

In 1959 Jacob and Wick\cite{Jacob:1959at} introduced the helicity formalism for the representations of the Poincare group.  Early papers on helicity contained detailed discussions of the Wigner rotations and phase conventions.  In 1966 Penrose and Newman\cite{NP} ($NP$) applied the helicity formalism to a problem in gravitational radiation.  For unknown reasons reference to Jacob and Wick, and any discussion of frame-fixing were omitted.   Instead a notion of ``spin-lowering'' was introduced. 

Spin-lowering is the addition of units of orbital angular momentum to the spin angular momentum to make objects of net helicity zero. The $NP$ operators $\eth$ operating on an object of helicity $s$ are defined by \ba  \eth =  sin^{s}\theta ( {\p \over \p \theta}  + {i csc \theta} {\p \over \p \phi} ) sin^{-s}\theta . \ea Lowering is done by $ \eth^{\dagger}$. The interior bracket is simply the ordinary orbital angular momentum raising and lowering operators $L_{x}\pm i L_{y}$ relative to a basis of $\hat x= \hat \theta, $, $\hat \phi=\hat y$.  Thus spin lowering (raising) subtracts (adds) a unit of $\hat r(\theta, \, \phi)\cdot \vec L$.  The factors of $sin^{s}\theta $ are Jacobians needed for operating on objects of helicity $s$ (called ``spin-weight'' by $NP$).  As above the orientation of the $(x, y)$ or $\hat \theta, \, \hat \phi$ spin-frame is a convention.  $NP$ remarked cryptically that ``$\eth$ is invariant under transformations that preserve the metric on the sphere''.  This is the particular phrase that hides the problem: The metric of humanly-defined frames has no pretense to invariance, and is not an observable.  

$NP$'s program continued with objects constructed by lowering or raising the spin-weight to zero and expansion in spherical harmonics.  Integration by parts clarifies this:  \ba  \bracket{lm}{\eth^{s}\rho_{s}} =\bracket{\eth^{s \dagger} \, lm}{\rho_{s}}.   \label{byparts}  \ea  The left hand side shows the raised or lowered states expanded in states of orbital angular momentum $l, m$.  The right hand side is the expansion of the original object in a basis defined by operating with $\eth^{s \dagger} $.   The basis of $\ket {\eth^{s \dagger} \, lm}$ were called the spin-weighted harmonics $_{s}Y_{m}^{l}$.  

It seems the $_{s}Y_{m}^{l}$ initially appeared to be something new and significant.  However Goldberg {\it et al}\cite{Goldberg:1966uu} soon showed that the expansion was simply an ordinary tensor spherical harmonics expansion in the helicity basis.  The identity needed is simply \ba  _{s}Y_{m}^{l}(\theta, \, \phi)=  { 2 l+1   \over 4 \pi  }^{1/2}D_{-sm}^{l}(\phi,\, \,\theta, \psi=0) .\ea  Notice that a convention-dependent zero-point of $\psi_{0} =0$ appears.  This is an $SO(2)$ tip of the iceberg of frame-fixing. By omitting any mention of the angle $\psi$ and the need for frame-fixing, $NP$ seem to have created a misleading impression, to whit: one can misinterpret Eq. \ref{byparts} as projection onto an {\it absolutely meaningful} basis free of conventions. Yet there exists an infinite number of equivalent bases.  The distinction of spin and orbital frames also underlies our use of index $j$ for total angular momentum, as opposed to the inappropriate symbol $l$ for orbital angular momentum in some absolute frame. 

{\it Electric and Magnetic Modes:} A complementary construction of tensor spherical harmonics was given by Zerillii in 1970.  Zerilli\cite{zerilli} re-introduced the terminology of ``electric'' and ``magnetic'' linear combinations that had been coined by Matthews\cite{matthews}.  Yet Zerilli failed to emphasize that the parity of the basis elements is simply a mathematical detail having nothing to do with the parity of objects expanded within it.  Thorne\cite{Thorne:1980ru}, for example, used both the electric and magnetic basis elements to expand gravitational radiation for a metric of even parity.  Similarly, in electrodynamics\cite{jackson} the vector spherical harmonics are confusingly named electric and magnetic while a general electric field (say) needs to be expanded in both electric and magnetic modes.  It should be clear that the conventions of   ``electric'' and ``magnetic'' basis elements represent nothing observable by themselves, and the freedom of frame-fixing of spin and orbital bases remains unchanged no matter how the elements are re-arranged. 

\subsubsection{Gradients and Curls} 
\label{sec:helm}

KKS avoid spin-lowering by contracting $\rho_{ij}$ with conventions for basis elements made from the derivatives of $Y_{m}^{l}$.  They demonstrate that their results are just the same as ZS due to an integration by parts (Eq. \ref{byparts}).  The hidden frame-fixing convention lies in using the derivatives defined via the oribital frame to make the spin frame, overlooking the Left-Right setup freedom. Obviously many other conventions exist, and the evident correlation of the spin-frames and orbital frames (oriented along the conventional polar axis) chooses a convention for Left-Right setup symmetry.  

The method of KKS makes the arbitrariness of frame fixing explicit, while it is complete hidden in the $NP$ method, generating a small paradox.  To support belief that the basis has an absolute meaning, KKS suggested the combinations of derivatives $\NN$ and $\hat r \times \NN$ should be viewed as decomposing the field into ``gradient'' ($G$) and ``curl'' ($C$) modes.  The authors cite the Helmholz theorem that such a decomposition should be unique.  

However there is no parallel of the flat space Helmholz theorem on the sphere.  The analogy fails even for the $s=1$ vector spherical harmonics.   Suppose $\vec v( \vec x)$ is a vector field on open flat 3-space, and one seeks $\phi, \, \vec A$ such that \ba     \vec v = \NN \phi + \NN \times \vec A .\ea  In flat space this is solved \ba \NN \cdot \vec v=  \nabla^{2} \phi ; \nn \\ \NN \times \vec v= -\nabla^{2} \vec A + \NN \NN \cdot \vec A ; \nn \\ \phi = {\NN \cdot \vec v \over \nabla^{2}}; \:\:\: \vec A = \vec v - \NN \phi . \ea The operator $1/\nabla^{2}$ is defined so that $$ \nabla^{2} ({1 \over \nabla^{2}})_{xx'}=\delta^{3}(\vec x-\vec x').  $$ Such Green functions have infinitely many symmetries $1/\nabla^{2} \ra 1/\nabla^{2}+ h$, where $\nabla^{2}h=0$ determines homogeneous solutions fixed by boundary conditions.  The usual Coulomb kernel is then fixed by choosing solutions that vanish at infinity.  Yet there is no analog of ``infinity'' on the sphere, and all solutions are doubly periodic.  Hence there is no unique definition of $1/\nabla_{T}^{2}$, and every effort to construct one creates a convention, which is the frame-fixing in yet another form.    

We construct and discuss the Green function $1/\NN_{T}^{2}$ in an Appendix. For an intuitive understanding of the breakdown of uniqueness, recall the physical interpretation of $1/\nabla_{T}^{2}$ is the potential of a point charge, with electrostatic field $-\NN_{T}/\nabla_{T}^{2}.$  Gauss' Law then requires the point charge be cancelled by another an equal and opposite charge somewhere on the sphere.  It is simple, but by no means mandatory, to place a canceling point charge on the opposite side of the sphere. This creates an axial anisotropy convention... indeed a new axis is introduced for each and every source!  The corresponding spin-frames are the unit vectors of the electric field from the initial charge to the arbitrary location of the mirror charge.  When the arbitrary mirror charge is re-located the arbitrary spin-frames are revised: and therein lies the frame-fixing. 

The failure to have an ordinary Helmholz theorem is topological and unavoidable.  It is interesting that this problem has been noticed in oceanography\cite{ocean}.  The authors of Ref. \cite{ocean} state ``The decomposition of an eddy flux into a divergent flux component and a rotational flux component is not
unique in a bounded or singly periodic domain...assertions made under the assumption of uniqueness, implicit or explicit, may be meaningless.'' We don't say the process is meaningless.  Conventions can be fixed to make it meaningful. 
However convention dependence must be brought out when comparing to observables. 
 
\section{Classifications} 

Here we review the classifications of the density matrix in our scheme, towards developing convention-free invariants.

\subsubsection{\it Classification of $\rho$}  Under the spin subgroup $\rho \sim 1 \otimes 1$ is reducible, and $1 \otimes 1 \ra 2 + 1 + 0$, where the integers stand for $S$.   Let $(S, \, s)$ stand for spin and helicity.  Then  $Q\pm iU \sim (2, \, \pm 2) $, $V \sim (1, 0) $, $I \sim (0,0) .$  Other components are absent because $\rho$ is transverse. Since we are using $\rho$ as a $3 \times 3$ matrix, the wave number $\vec k$ is the 0 eigenvector of $\rho$, oriented along $-\hat r$. 

Under spin-parity the electric field's vector $\vec E \ra -\vec E$, so that $\rho \ra \rho$ is even. The spin-1 associated with $V$ is a pseudovector because the Clebsch for $1 \times 1 \ra 1$, namely $\epsilon_{ijk}$, needs a handedness convention.  From products of $\rho$ and $\hat k \sim(1, 0)$, and using the addition rules of angular momentum, there is one pseudoscalar, Stokes $V =\epsilon_{ijk}\rho_{ij}\hat k_{k}.$  From $\rho \otimes \rho$ we can make one new scalar beyond Stokes $I$, the degree of polarization $\sqrt{ Q^{2}+U^{2}}.$  
This finishes the local classification of $\rho$.

\subsubsection{The Data Map} 

Given a set of measured angles in data $\phi_{i} \, \theta_{i} \, \psi_{i}$, the empirical distribution is a sum of delta functions concentrated on data values. The empirical distribution $f_{data}$ is expanded in the basis of $D$'s as follows, \ba f(\phi_{i} \, \theta_{i} \, \psi_{i}) =   {  2j+1  \over 8 \pi^{2}    }   \sum_{i} \, \rho_{sm}^{j}D_{sm}^{j}; \nn \\  \rho_{sm}^{j}= \sum_{i} \, D_{sm}^{j*}(\phi_{i} \, \theta_{i} \, \psi_{i}). \label{dmap} \ea Procedures for ``cleaning'' or interpolating data may appear here to incorporate features of instrumentation or to improve statistical reliability.  Whatever method is used, 
we call Eq. \ref{dmap} the {\it data map} from the angles to the coefficients. 

\paragraph{\it Frames from data:}  Interestingly, data summarized by $\rho_{sm}^{j}$ contains within itself certain preferred spin and orbital frames. Statistical analysis of the different frames may reveal new insights into the CMB, just as the orientation of the ordinary CMB multipoles has led to new information. 

The ``canonical frames'' are found by diagonalizing $\rho_{sm}^{j}$ for each $l$, writing \ba \rho_{sm}^{j} = u_{s}^{j \a} \lambda^{j\a}(v^{jT})_{m}^{\a}, \label{svd} \ea where $T$ denotes the transpose.  In coordinate-free notation \ba  \rho^{j} = \sum_{\a} \ket{u^{j\a}}\lambda^{j\a} \bra{v^{j\a}} \label{svd2} .\ea This is the singular value decomposition ($svd$), which is unique. Here $\Lambda^{\a j }= \lambda^{(1)j }, \, \lambda^{(2)j }$ are the singular values (``eigenvalues''), invariant under $SO_{spin}(3) \times SO_{orbital}(3)$, and therefore invariant under the full rotation group. Matrices $u_{s}^{j \a} $, $v _{m}^{j\a}$ are the orthogonal frames constructed by diagonalizing the Hermitian combinations $\rho^{j} \rho^{j\dagger}$ and $\rho^{j\dagger}\rho^{j}$: \ba \rho^{j} \rho^{j\dagger}_{ss'} = (\,  u^{j} (\Lambda^{j})^{2} u^{Tj} \,)_{ss'} ; \nn \\  \rho^{j\dagger}\rho^{j}_{mm'} =  ( \,  v^{j} (\Lambda^{ j})^{2} v^{Tj} \,)_{mm'}. \ea  For CMB density matrices classified under $s=0, 1, 2$ there are at most two orthogonal terms, $v _{m}^{(1)}$, $v _{m}^{(2)}$ for each $j$, while the range of $m=0, \, 1, \, 2, \, 3 ...$ is unrestricted.  As eigenvalues of a positive Hermitian matrix, the diagonal elements $(\Lambda ^{j})^{2}$ are real and positive, and $\Lambda ^{j}$ are conventionally defined to be positive to fix the phases of $u_{s}^{j \a} $ and $v_{m}^{j \a} $. 

The diagonal direct products show that any data for $\rho_{sm}^{j}$ can be interpreted for each $j$ as two true invariants, plus inherent orientation of spin frames strictly correlated with inherent oriented orbital frames. The orbital frames of course contain more information than the invariants: they are {\it covariant} under rotations.   The invariants are readily extracted from ordinary traces of the matrices:  \ba   ( \lambda^{(1)j })^{2} +  ( \lambda^{(2)j })^{2}= tr(\rho^{j} \rho^{j\dagger});  \nn \\  (\lambda^{(1)j })^{2} ( \lambda^{(2)j })^{2}=det(\rho^{j} \rho^{j\dagger}),   \ea where $tr$ and $det$ denote the trace and determinant.  These are equivalent to the invariants exhibited earlier.  

\subsubsection{\it Intrinsic Frames of Basis Elements:} 

Just as above, the convention-dependence of the basis states can always be discovered by the $svd$ procedure.  Given elements $D_{sm}^{j}$ in a basis labeled ``$\nu$'', there is a unique expansion \ba  D_{sm}^{j}(\nu) = u_{s}^{j \a}(\nu)  \lambda^{j\a}(v^{j\a})_{m}^{\dagger}(\nu) . \ea The particular orthogonal matrices encode the net history of parallel transport used to construct the particular basis, which ``latches onto'' unit vectors $(v^{j\a})_{m}$ and ``returns'' $u_{s}^{j \a}$ as a result.  

Hence if one experimenter generates a set of $D_{sm}^{j}$ matrices according to a given convention, another experimenter can find the intrinsic frames, which are equivalent to the ordering and phase conventions.

\section{Distributions} 

We now turn to the construction of manifestly invariant distributions of the density matrix. 
An invariant distribution must be a function of an invariant argument $f=f(\rho, \, n) $, where
$n$ represents tensor parameters of appropriate type. When cast into our basis, $f=f( \rho_{sm}^{j}, \, n_{sm}^{j})$, where $n_{sm}^{j}$ are the matrix elements of parameter $p$.  It is useful that each subspace of index $j$ is independent.  With  distributions one can examine various correlations defined as moments, or use other tests such as likelihood, depending on the complexity of description sought.  In order of decreasing symmetry, we list various cases useful to categorize data. 

\subsubsection{Isotropy and Parity} 

Isotropy requires $f(\rho, \, p) \ra f(\rho) =f( I, \, Q^{2}+U^{2}, \, V)$, since these arguments are the only invariants made from $\rho$.  If the distribution is invariant under parity, then $f( I, \, Q^{2}+V^{2}, \, V)=f( I, \, Q^{2}+U^{2}\, -V)$, yielding $<V>=0$, and similarly for odd powers of $V$. 

We have shown that $E$ and $B$ modes are not true invariant classes.  There comes a paradox that parity symmetry has been used to make predictions of $E$ and $B$ mode correlations.  Examine the bilinear correlation \ba   C_{ss', \, mm'}^{ j}= <    \rho_{sm}^{j} \rho_{s'm'}^{j*}>,\ea where $*$ indicates complex conjugation.  Isotropy requires that $C_{ss', \, mm'}^{ j}$ be an invariant tensor. 

The only rotationally invariant tensor symmetric in $ss'$ and $mm'$, with generally different dimensions, is given by  \ba C_{ss', \, mm'}^{j}= C^{j} \, \delta_{ss'}\delta_{mm'}  \label{iso}.\ea   We then find the correlation \ba < \rho_{m}^{E, j} \rho_{m}^{B, j*} > = C^{j}\delta_{mm'}( \delta_{22}- \delta_{-2-2}) =0. \label{zero}\ea That is, we find \ba <\rho_{m}^{E, j} \rho_{m}^{B, j*} > =0. \ea  This a consequence of {\it isotropy} without mention of {\it parity symmetry}.  In the current literature Eq. \ref{zero} is identified as a consequence of $\pi_{s}$ symmetry, not true parity.  Perhaps $\pi_{s}$-invariance might be something reasonable to propose, but it lacks the fundamental character of true parity.  It does not break parity to find $<\rho_{m}^{E, j} \rho_{m}^{B, j*} >  \neq 0$ since the actual CMB is not isotropic.  

\subsubsection{Axial Dependence} 

The next simplest symmetry class would contain axial parameters $\hat n$, which might be vectors (oriented) or the eigenvectors of tensors (unoriented).  There is an orbital scalar made linearly via $\hat r \cdot \hat n$.  One can make a spin-pseudoscalar $\epsilon_{ijk}\hat n_{i}\rho_{ij}$ which depends on Stokes $V$, and a quadratic
spin-scalar by the combinations $\rho_{ij} \hat n_{i} \hat n_{j}$.  Continuing, from a single $\hat n$ there is an infinite number of tensors $\hat n \otimes \hat n ...\hat n$ of arbitrary order, which can be reverted to objects transforming like helicity $s$ or orbital angular momentum $m$: finally tensors $n_{sm}^{j}$ in each subspace $j$.  Current data for the CMB does have a symmetry-breaking axis, obtained from the dipole ($j=1$) intensity map.  Several authors\cite{cmbaxis} have noted that a nearly coincident axis appears for $j=2, 3, 4$. Our analysis suggests that an axial dependence might be related to the $<TE>$ correlation WMAP observes\cite{Kogut:2003et}. 

When isotropy fails, Eq. \ref{iso} fails.  It is clear that correlations of the type $<    \rho_{sm}^{j} \rho_{s'm'}^{j*}>$ are not manifestly invariant.  The rotationally invariant quadratic correlations are of the form \ba C^{j} = \sum_{sm} <    \rho_{sm}^{j} \rho_{sm}^{j*}>. \ea The summands are linear combinations of the invariants listed earlier.  By comparing this invariant with $C_{ss', \, mm'}^{ j}$ various test of isotropy can be undertaken. 

Now suppose some signal violating isotropy is observed:  how does one proceed?  By analogy with familiar Gaussian correlations, the $C_{ss', \, mm'}^{ j}$ serve as model parameters one can use to build up heuristic distributions.  Thus one can made simple likelihood tests, comparing $f=f(\rho)$ (the isotropic default) with $f=f( \rho_{sm}^{j}C_{ss', \, mm'}^{ j}\rho_{s'm'}^{j*})$, which uses an invariant built from the data itself. (We hasten to add that data sets are enormous and complex, and data analysis itself is an art.)

We note that in the presence of anisotropy the relative orientation of the space and spin-frames can obviously be fixed in an absolutely meaningful way.  In Cartesian language, we may define $Q_{n} \sim \hat n_{i}\rho_{ij} \hat n_{j};  \,  U_{n} \sim \hat n_{i}\rho_{ij}   \epsilon_{jpq} \hat n_{p} \hat k_{q}. $  In this case $Q$ and $U$ can be cast into physically-based coordinates, not from $\rho$ alone, but from $\rho$ relative to $\hat n$. Now with $U_{n}$ a pseudoscalar, parity symmetry {\it with the anisotropic distribution} will predict $<Q_{n}U_{n}>=0$, a meaningful relation for $E$ and $B$ modes defined relative to $\hat n$.  This explains why cartoons of $E$ and $B$ modes have visible parity properties: they are cartoons emphasizing a particular $\hat n$ axis.   

\subsubsection{  $\vec L \cdot \vec S$ Coupling} 

The most familiar coupling of space- and spin-angular momentum in quantum mechanics is called the ``$\vec L \cdot \vec S$'' coupling.  The analogue for CMB data is straightforward to construct.  Let $\vec S_{ss'}^{(j)}$ be matrix elements of the spin- angular momentum generators. Only the $s, \, s' =\pm 2$ components will actually be used.  Let $\vec L_{m'm}^{(j)}$ be matrix elements of the spin-$j$ generators - these are not gradient operators, but matrices of numbers.  Then the following is a dependent scalar for each $j$, \ba C_{LS}^{j} =\sum_{k} \,  tr(\,  \rho^{j\dagger} \vec S^{(j)} \cdot  \rho^{j} \vec L^{(j_{}} \  \,),  \nn \\ = \sum_{mm'ss'k}  \rho_{ms}^{* j}S_{ss'}^{(j) k }\rho_{s' m'}L_{m'm}^{(j)k} \ea   This represents a correlation of the global polarization and angle dependence on the dome of the sky.  The expected value is zero when there is isotropy in {\it either} the spin or space dependence.

\subsubsection{WMAP}
 
The observation of correlations that are convention-dependent is by no means useless information.  Results so far reported by WMAP\cite{Kogut:2003et} include a ``$<TE>^{l}$'' correlation showing surprising structure at small $l$.  Sometimes the correlation is defined by $$C(\theta) ={ \sum_{ij}Q'(\hat r_{Q, i})I(\hat r_{I, j} ) w_{i}w_{j} \over \sum_{ij}w_{i}w_{j}} , $$ where $\hat r_{Qi}$ and $\hat r_{Ij}'$ are location where $Q$ and $I$ are measured separated by angle $\theta$, and symbols $w_{i}, \, w_{j}$ are weights introduced to correct the data.  The modified Stokes $Q'$ is defined as the Stokes $Q$ parameter relative to an origin oriented along great circles connecting pairs of points $i, \,j$ 

One can understand WMAP's procedure as choosing spin-frames at $\hat r_{Q, \,i}$ via a fiducial $\hat z$ axis at $\hat r_{I,  \, j}$.  Fix index $j$; the great circles are simply the coordinate lines of local $\hat \theta_{I, j}$.  With this convention calculate the unvarnished $E$ modes of Eq. \ref {elecmag} relative to $\hat \theta_{I, j}$.  Continue, moving to the next $j$ and sum.  Does this make a convention-free observable?

Unfortunately no.  From our analysis of frame-fixing there is one free choice of frame orientation to fix per $l$. This convention is a convention however it is done.  For instance one might choose to measure $Q'$ relative to the normals to the geodesic; then angle $\psi' \ra \pi/2+\psi'$, whereupon $<TE>^{l} \ra  - <TE>^{l}$. We do not believe that the sign of a physical observable is an arbitrary thing $l$ by $l$.  More directly it is just arbitrary to develop the frames by parallel transport from $\hat r_{I, j}$. (Suppose for instance that all the data came from a small region of the sky, as in earlier studies not covering the full sky.  Then a bias relative to this choice of $\hat z$ is obvious.)  We have shown no absolute frame fixing exists, and that $<ET>$ correlations are not true scalars. 

As mentioned earlier, comparison of theory and data treated within any fixed and consistent procedure can always be made.  WMAP interprets their correlation data as signals of a late ionizing phase, citing several theoretical predictions. We are unable to assess consistency of data analysis and theory which is liable to be quite complicated.  Yet breakdown of true invariance gives no immediate reason to doubt the consistency of the comparison reported.   It would certainly be interesting to explore the hidden freedom of convention when comparing data and models. 

It would also be very interesting to see more varied and versatile CMB polarization data analysis to complement more familiar power spectra. Testing distributions based on the invariant and covariant decompositions suggested earlier may have a rich potential for discovery. 

\medskip 

Acknowledgments:  A report on this work was given at {\it Miami 2005}, for which we thank the organizers.  Research partly supported under DOE grant number DE-FG02-04ER14308.

\section{Appendix}

In this Appendix we show analytically there exists no analog of the Helmholz theorem on the sphere.

We may show this by contradiction.  Consider a region ``1'' where $\NN_{T} \cdot \vec v_{T1} \neq 0$, where $T$ represents the transverse components in the tangent plane.  Use Gauss' Law to integrate over the region, $\int_{1} d^{2}x_{T} \, \NN_{T} \cdot \vec v_{T1} \ra \oint_{1} d\vec S_{T} \cdot \vec v_{T1}. $ The same contour is the oppositely-oriented boundary of the complementary region ``2'' of the sphere.  Hence \ba  \oint_{1} d\vec S_{T} \cdot \vec v_{T1}.+ \oint_{2} d\vec S_{T} \cdot \vec v_{T2} =0, \ea meaning that the integrated divergence vanishes.  

We now seek an operator $1\nabla_{T}^{2}$ such that we recover the sole divergence of region 1, $$\nabla_{T}^{2} (  {  \NN_{T} \cdot \vec v_{T1}\over  \nabla_{T}^{2}}) =   (\NN_{T} \cdot \vec v_{T1}) .  $$ This is a contradiction, because the net divergence must be zero, while the net divergence in region 1 is non-zero.  The integration over region 1 will instead {\it assign} a corresponding anti-divergence to region 2.  What rule shall be used?  Without actually integrating over region 2 (the region 1 integration probing only the boundary) it is impossible to recover the correct divergence coming from region 2.  

A specific example: consider the divergence singularity of $\vec v \ra \hat \theta$ in the northern hemisphere.  The only possible scheme to re-distribute this singularity in a rotationally symmetric way must involve a constant divergence.  This defines a Green function $G_{0}$ such that \ba  \nabla_{T}^{2}G_{0}(\theta, \, \theta' ,\,\phi,\, \phi') = \delta(cos \theta-cos\theta')\delta(  \phi-\phi') -{1 \over 4 \pi}. \nn \ea   The integral of the right-hand side is zero, as required, but we have an operator which differs {\it everywhere} from the desired Green function.  It is clearly possible to force the canceling divergence to be an isolated singularity of opposite ``charge,'' but where on the sphere shall it be placed?  Thus all Green functions contain arbitrary symmetry-breaking conventions: introducing an artificial {\it axial anisotropy} is the most symmetric result possible.

 Construction of the Green functions goes as follows:  Define $G(\hat r, \, \hat r')$ such that on the sphere \ba \nabla_{T}^{2} G(\hat r, \, \hat r')=\delta^{2}( \hat r, \, \hat r'). \ea  With $\hat r(\theta, \, \phi)$ in a reference coordinate system, expand $G$ in spherical harmonics, whereby $\nabla_{T}^{2}  \ra l(l+1)$ on each mode.  Completeness suggests the usual solution dividing out the operator, namely  \ba  G_{0}= \sum_{m, \,  l=1}^{\infty} \,  {  Y_{m}^{l}(\theta, \, \phi)Y_{m}^{l*}(\theta', \, \phi') \over l(l+1)}  . \ea  Note the $l=0$ term is divergent and had to be omitted.  It follows \ba \NN_{T}^{2}G_{0}= \delta(cos\theta -cos \theta') \delta(\phi - \phi') -{1 \over 4 \pi}, \ea which smears an anti-charge uniformly over the sphere.  To compensate add any solution $h$ such that $\nabla_{T}^{2}h=1/4 \pi$.  Proceed by expansion, \ba  \nabla_{T}^{2}h=\sum_{lm}\, l(l+1) \, h_{lm} \, Y_{m}^{l}(\theta, \, \phi)Y_{m}^{l *}(\theta', \, \phi') ={1 \over 4 \pi} .\ea  Note that either the expansion breaks rotational symmetry with some non-trivial tensor $h_{lm}$, or $h_{lm}\ra h_{l}$ does not depend on $m$.  In the second case the addition theorem for spherical harmonics gives \ba \sum_{lm}\, l(l+1) \, h_{l} \, Y_{m}^{l}(\theta, \, \phi)Y_{m}^{l*}(\theta', \, \phi')= \nn \\ = \sum_{l} \,{ (2 l+1)  l(l+1) \, h_{l} \over 4 \pi} P_{l}(cos \gamma) ={1 \over 4 \pi}, \ea where $\gamma$ is the angle between $\hat r$ and $\hat r'$ and $P_{l}$ are the Legendre polynomials. Comparing the left and right side, $$ h_{l} ={ \delta_{l0} \over (2 l+1)  l(l+1)  }, $$ meaning that $h_{l}$ does not exist to satisfy the requirements.   This leaves the first case, $h_{lm} \neq 0$.  An infinite number of solutions exist but all break rotational symmetry.  For example $\nabla_{T}^{2}(Y_{m}^{l} -Y_{0}^{l}) =0$ for all $l$ and any $m$.  
 
We can apply this calculation to the discussion of charge-and anti-charge of Section \ref{sec:helm}.  The ansatz  is \ba  G_{1}= \sum_{m, \,  l=0}^{\infty} \,  {  Y_{m}^{l}(\theta, \, \phi)Y_{m}^{l*}(\theta', \, \phi') \over l(l+1)}-{  Y_{m}^{l}(\pi-\theta , \, -\phi)Y_{m}^{l*}(\theta', \,  \phi')  \over l(l+1)} , \ea whose Laplacian develops two $\delta$ function singularities with the offensive $1/4 \pi$ terms canceling.


\begin{thebibliography}{}

\bibitem{Batschelet} E. Batschelet, {\it Circular statistics in biology}, (Academic Press, London, 1981). 

\bibitem{KKS}  M.~Kamionkowski, A.~Kosowsky and A.~Stebbins,
  Phys.\ Rev.\ D {\bf 55}, 7368 (1997).

\bibitem{ZS}    M.~Zaldarriaga and U.~Seljak,
  Phys.\ Rev.\ D {\bf 55}, 1830 (1997).

\bibitem{sakurai} J. J. Sakurai, {\it Modern Quantum Mechanics} (Addison-Wesley, 1994). 

\bibitem{NP} E.~T.~Newman and R. Penrose,   J.\ Math.\ Phys.\ \bf{7}, 863 (1966) 

\bibitem{Goldberg:1966uu}
  J.~N.~Goldberg, A.~J.~MacFarlane, E.~T.~Newman, F.~Rohrlich and E.~C.~G.~Sudarshan,
  J.\ Math.\ Phys.\  {\bf 8}, 2155 (1967).

\bibitem{zerilli} F. J. Zerilli,  J.\ Math.\ Phys.\ \bf{11}, 2203 (1963) 


\bibitem{matthews} J. Matthews, J. Soc. Ind . Apl. Math, \bf{10}, 762, (1962). 
\bibitem{Thorne:1980ru}
  K.~S.~Thorne,
  Rev.\ Mod.\ Phys.\  {\bf 52}, 299 (1980).

 \bibitem{jackson} {\it Classical Electrodynamics}, by J. D. Jackson, 2nd Edition, (Wliey, New York, 1975).

\bibitem{wigner} E. P. Wigner, {\it Group Theory} (Academic Press, New York, 1959).


\bibitem{Jacob:1959at}
  M.~Jacob and G.~C.~Wick,
  Annals Phys.\  {\bf 7}, 404 (1959)
  [Annals Phys.\  {\bf 281}, 774 (2000)]




\bibitem{ocean} B. Fox-Kemper,R. Ferrari, and J. Pedlowski, Jour.Phys.  Ocean. \bf{ 33}, 478 (2002) 

\bibitem{cmbaxis}  A.~de Oliveira-Costa, M.~Tegmark, M.~Zaldarriaga and A.~Hamilton,
  Phys.\ Rev.\ D {\bf 69}, 063516 (2004); J.~P.~Ralston and P.~Jain,
  Int.\ J.\ Mod.\ Phys.\ D {\bf 13}, 1857 (2004)
 




\bibitem{Kogut:2003et}
  A.~Kogut {\it et al.},
  Astrophys.\ J.\ Suppl.\  {\bf 148}, 161 (2003);   L.~Page {\it et al.},
  arXiv:astro-ph/0603450.




\end{thebibliography}
\end{document}